\documentclass[twocolumn]{svjour3}          

\smartqed  
\usepackage{graphicx}
\journalname{Journal of Real-Time Image Processing}
\begin{document}

\title{Architecture-Aware Optimization of an HEVC decoder on Asymmetric Multicore Processors
}


\author{Rafael Rodr\'{i}guez-S\'{a}nchez         \and
        Enrique S. Quintana-Ort\'{i} 
}


\institute{Rafael Rodr\'{i}guez-S\'{a}nchez \and  Enrique S. Quintana-Ort\'{i} \at
              Depto. Ingenier\'{i}a y Ciencia de Computadores \\
              Universidad Jaume I \\
              Avda. Sos Baynat s/n \\
              12071, Castell\'{o}n de la plana, Spain \\
              \email{\{rarodrig,quintana\}@uji.es}           
}

\date{Received: date / Revised: date}

\maketitle

\begin{abstract}
Low-power asymmetric multicore processors (AMPs) attract considerable attention due to their 
appealing performance-power ratio for energy-constrained environments. 
However, these processors pose a significant programming challenge due to the
integration of cores with different performance capabilities, asking for an asymmetry-aware scheduling 
solution that carefully distributes the workload. 

The recent HEVC standard, which offers several high-level parallelization 
strategies, is an important application that can benefit from an implementation tailored for the low-power AMPs present in 
many current mobile or hand-held devices. In this scenario, we present an architecture-aware implementation of an HEVC decoder 
that embeds a criticality-aware scheduling strategy
tuned for a Samsung Exynos 5422 system-on-chip furnished with an ARM\textregistered\ big.LITTLE\texttrademark\ AMP. 
The performance and energy efficiency of our solution
is further enhanced by exploiting the NEON\texttrademark\ vector engine available in the 
ARM big.LITTLE architecture. Experimental results expose a 1080p real-time HEVC decoding 
at 24 frames/sec, and a reduction of energy consumption over 20\%.

\keywords{
HEVC \and
asymmetric multicore processors \and
scheduling \and
vector intrinsics \and
real-time decoding \and
energy efficiency}
\end{abstract}

\section{Introduction}
\label{intro}

High Efficiency Video Coding (HEVC)~\cite{HEVC} 
is the successor of the most widely-used video codec, H.264/Advance Video Coding (AVC)~\cite{H264} and, therefore, a serious
candidate to become the state-of-the-art tool for video compression in the near future.
One crucial requirement for video compression, which shaped the HEVC standard, is adaptability, especially 
for practical consumer electronics applications. 
In particular, video content should be preferably distributed in a format that is in accordance 
with the display and memory capabilities, processing power and computational 
constraints of consumer electronics appliances as well as with the network bandwidth. 

The Joint Collaborative Team on Video Coding (JCT-VC), which includes video experts from both the ISO/IEC 
Moving Expert Group (MPEG) and the ITU-T Video Coding Expert Group (VCEG),
designed the new HEVC standard taking into account these requirements.
Concretely, HEVC was conceived to deliver high quality multimedia services, 
with reasonable efficiency and performance over bandwidth-constrained networks. In addition,
a major concern during the standardization process was to minimize the computational requirements and energy consumption~\cite{HEVC3}. 
Although several coding tools were not included in HEVC, due to their complexity, 
in comparison with its predecessor H.264/AVC~\cite{H264}, 
the computing and energy demands of HEVC have been considerably increased~\cite{Rodriguez14}.

Asymmetric multicore processors (AMPs) have been recently proposed for severely
energy-constrained environments, especially for mobile appliances, where heterogeneity in applications is mainstream.
These architectures integrate two (or more) types of cores with different capabilities, 
which share the instruction set architecture but differ 
in micro-architecture. A practical example is the ARM\textregistered\ big.LITTLE\texttrademark\ 
AMP included in Samsung's Exynos 5422 system-on-chip (SoC),
which features both ARM Cortex-A15\texttrademark\ and Cortex-A7\texttrademark\ cores. 
The former type of core delivers higher performance than the Cortex-A7 counterpart, at the expense of higher power dissipation
rate, while the Cortex-A7 cores can potentially deliver a more favourable perfor\-mance-power ratio.

In this paper, we integrate an asymmetry- as well as criti\-cality-aware 
scheduling strategy into the multi-threaded implementation of the 
libde265 library specifically adapted for the Exynos 5422 SoC. 
We will target the wavefront parallel processing (WPP) scheme~\cite{WPP} defined for HEVC, in which multiple
``regions'' of a single frame can be processed simultaneously. Compared with other alternatives, such as the
tiling and slicing approaches, WPP does not limit intra-prediction nor resets CABAC probabilities, 
avoiding a reduction of the encoding efficiency. However, as we will show later, this parallelization
strategy posses a considerable challenge for an AMP, due to the dependencies when reconstructing the CTU rows.

To exploit the asymmetric hardware concurrency,
our parallel solution migrates the threads taking into account the 
dependencies of the tasks in execution, so that those tasks with higher priority are always executed in
the fast (big) ARM Cortex-A15 cores, while the non-priority tasks run on the slow (LITTLE) Cortex-A7 cores. 
Thus, the key to this strategy is a policy that moves tasks between fast and slow cores on-the-fly as the tasks' priorities vary. 
The implementation is further enhanced by including ARM-specific NEON\texttrademark\ optimizations into certain modules of the library,
a capability that is not available in the reference HEVC library.
The experimental evaluation of our proposal on an ODROID-XU3 board, furnished with an Exynos 522 SoC, demonstrates
the advantage of the new decoder in 
terms of decoded frames per second (FPS), but also from the perspective of energy efficiency.

The rest of the paper is structured as follows. 
In Section~\ref{related}, we briefly describe the HEVC standard and discuss some
related works.
In Section~\ref{deco}, we experimentally 
illustrate the poor (computational) performance and energy efficiency of the reference multi-threaded implementation of the libde265
library on the target ODROID-XU3 board/Exynos 5422 SoC.
In Section~\ref{proposal}, we introduce our strategies to adapt the original libde265 
implementation to the asymmetric ARM big.LITTLE architecture; and in Section~\ref{performance} we report 
the performance and energy-efficiency results of the new solution.
Finally, Section~\ref{conclu} closes the paper with a few concluding remarks.

\section{Technical Background}
\label{related}

\subsection{The HEVC standard}

HEVC \cite{HEVC} introduces new coding tools with respect to its predecessor, H.264/AVC, 
as well as improves upon alternative previous encoders. 
The HEVC standard can reach the same subjective video quality 
than its predecessor by using half the bit-rate \cite{HEVC2,HEVC3}, while
notably increasing coding efficiency. 

\begin{figure}[t]
\begin{minipage}[b]{1.0\linewidth}
  \centering
  \centerline{\includegraphics[width=\linewidth]{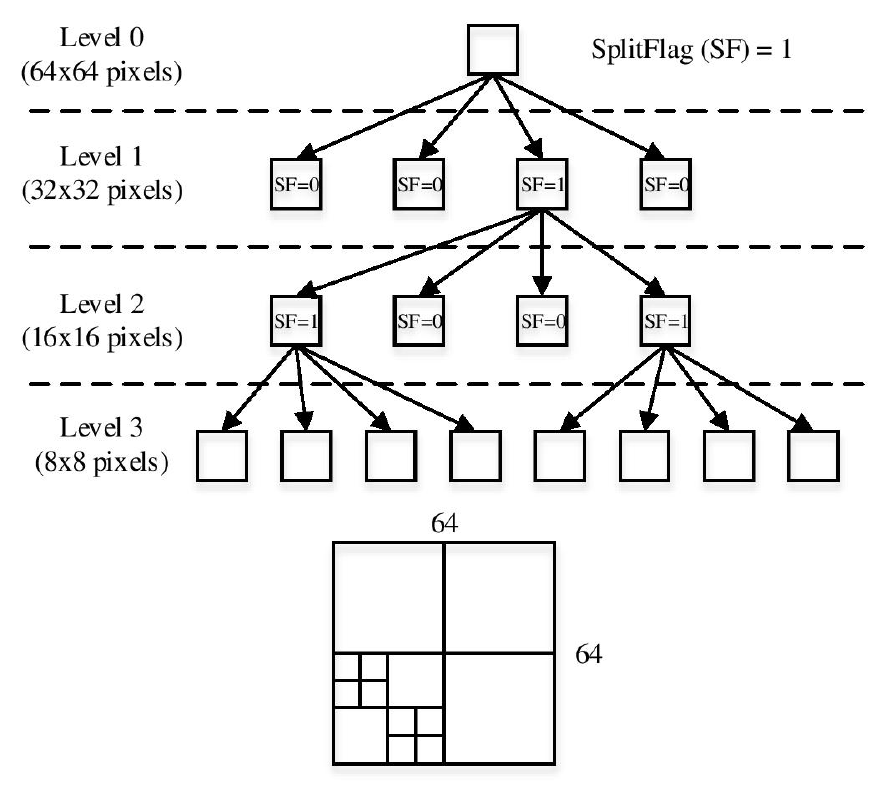}}
\end{minipage}
\caption{HEVC image partitioning.}
\label{fig:HEVC}
\end{figure}

One important change in HEVC affects image (frame) partitioning. 
The new standard introduces three new 
concepts: Coding Unit (CU), Prediction Unit (PU), and Transform Unit (TU). A frame is partitioned 
into Coding Tree Units (CTUs), each containing a single luma Coding Tree Block (CTB) and two chroma 
CTB blocks. Each CTU is further partitioned into several square regions, of variable size, called CUs. 
Each CU, consisting of 8$\times$8 to 64$\times$64 pixels, may contain one or several PUs 
and TUs. This partitioning can be performed within each sub-area recursively, until it has a size 
of 8$\times$8 pixels; see \figurename~\ref{fig:HEVC}. At PU-level, intra- and inter-predictions are 
carried out with sizes ranging from 4$\times$4 to 64$\times$64 pixels. The CUs may be further split into TUs, 
which contain the coefficients for transformation and quantization in the form of square blocks of pixels.
This structure leads to a more flexible coding that suits the particularities of the frame.

Tiles~\cite{Tiles} and WPP are two new technologies 
introduced in the HEVC standard to support high-level parallelism, in both cases by allowing the division of frames 
for parallel decoding. Additionally, HEVC inherits the slice concept from previous standards. 
When working by tiles, a frame can be divided into rectangular groups of CTUs, which are treated as independent 
decoding tasks. Alternatively, with WPP, CTU rows can be decoded in parallel though,
due to data dependencies, the decoding of a CTU row must be proceed with a delay of two CTUs with 
respect to that of the previous CTU row. This yields a wavefront parallel processing scheme that names this approach.

\subsection{Related work}

In the 
literature we can find general strategies to accelerate HEVC video decoders, 
but also proposals for energy-constrained devices which pursue similar goals to those set for our work.

In~\cite{Chi15}, the authors use SIMD instructions to accelerate all major modules of an 
HEVC decoder, obtaining speed-ups of up to $5\times$ on mobile and desktop platforms, to deliver 1080p real-time 
decoding. Similarly, the authors of~\cite{Duan14,Meng14,Bariani14} report 720p real-time decoding 
on an ARM Cortex-A9 platform and 1080p real-time decoding on an Intel platform.
Although all these works leverage SIMD optimizations on energy-constrained devices, 
none of them analyzes the impact of this type of optimizations in terms 
of energy consumption. In~\cite{chi13} the authors introduce a methodology to deal with the dependencies intrisinc to WPP. 
Concretely, they adapt the WPP approach and process several partitions as well as several pictures in parallel. This presents
the advantage of maintaining the amount of (thread-level) parallelism during the execution. This approach is 
further optimized in~\cite{chi12} and compared with other high-level parallelization
strategies defined in the HEVC standard.

Hardware implementations of the HEVC decoder are also a hot research topic. 
In~\cite{Tikekar14}, a 40nm CMOS hardware decoder is introduced that
can process 4k videos in real-time. In~\cite{Ju14} a more energy efficient 28nm CMOS hardware
decoder is presented. Moreover, there are also proposals that focus on a concrete 
module of the HEVC decoder such as~\cite{Chiang13,Zhu13,Kalali12}.

In the context of mobile and hand-held devices, the authors of~\cite{Raffin15,Raffin15-2} describe several strategies for power 
optimization of a real-time software HEVC decoder on a NEON architecture. These strategies exploit 
data-parallelism, task-parallelism, and dynamic vol\-tage-fre\-quency scaling (DVFS); however, only one type of core 
in an ARM big.LITTLE AMP is used at a time. In~\cite{Nogues14,Nogues15}, the authors reduce the 
filtering complexity to diminish energy consumption at the expense of a significant degradation in the 
subjective video quality. 
Energy reductions of up to 28\% are reported there for an ARM bit.LITTLE core, but 
the authors only use one type of core. Some complementary work also aims to 
accelerate a concrete module on these energy-constrained devices~\cite{Yong13}.

Additionally, we can find some efforts to characterize the execution time and 
energy consumption 
of an HEVC decoder; for example,
complexity-related aspects in the standardization
process~\cite{Bossen12}; 
energy consumption of multicore CPUs and hardware-accelerated decoders~\cite{Benmoussa15};
exploiting \textit{race to idle} and \textit{slack} via DFVS for energy efficiency~\cite{chi14}; etc.

\section{Multi-Threaded Implementation of the HEVC Decoder}
\label{deco}

Our proposal for HEVC decoding builds upon the open source libde265 library. 
This implementation is written in C++ from scratch, and provides a plain C API that enables 
its integration into other software. At present, there are 
many applications that rely on libde265 for HEVC video decoding, and 
there exist plugins for gstre\-amer, VLC, Windows DirectShow, 
and the ffmpeg decoder, among others.

The libde265 library includes novel tools to support the high-level parallel schemes defined in the HEVC standard, 
in particular tile-based and WPP multithreading. 
Additionally, 
the library integrates Intel's Streaming SIMD Extensions (SSE) 
to further enhance performance on x86-based architectures. 

This reference library follows a master-slave approach where a single master thread creates and enqueues tasks, 
while several (worker) threads process these tasks. Given a frame, in WPP
the master thread initially creates one task per CTU row, 
and keeps track of task dependencies. 
In particular, if we number the CTUs in the frame from the top left-corner to the bottom-right one starting at $(0,0)$,
where the first index denotes the CTU row, 
then CTU $(i,j)$ can not be reconstructed until CTU
$(i-1,j+2)$ has been completely processed.
(This constraint does not apply when the parallelization approach is based on slices or tiles.) 
Once all CTU rows have been reconstructed (by the worker threads),
in a second step, the tasks corresponding to the Deblocking and SAO filters are generated and these filers are applied. 
The library generates three tasks per CTU row: one to filter the vertical edges, one to filter the 
horizontal edges, and one for the SAO filter. In this second step, a synchronization 
of the tasks
is always needed since, in order to perform the vertical filtering of a concrete CTU, up to three CTUs must be 
previously horizontally filtered. The same applies to the SAO filter.


\subsection{Experimental setup}
\begin{figure}[t]
\begin{center}
\includegraphics[width=\columnwidth]{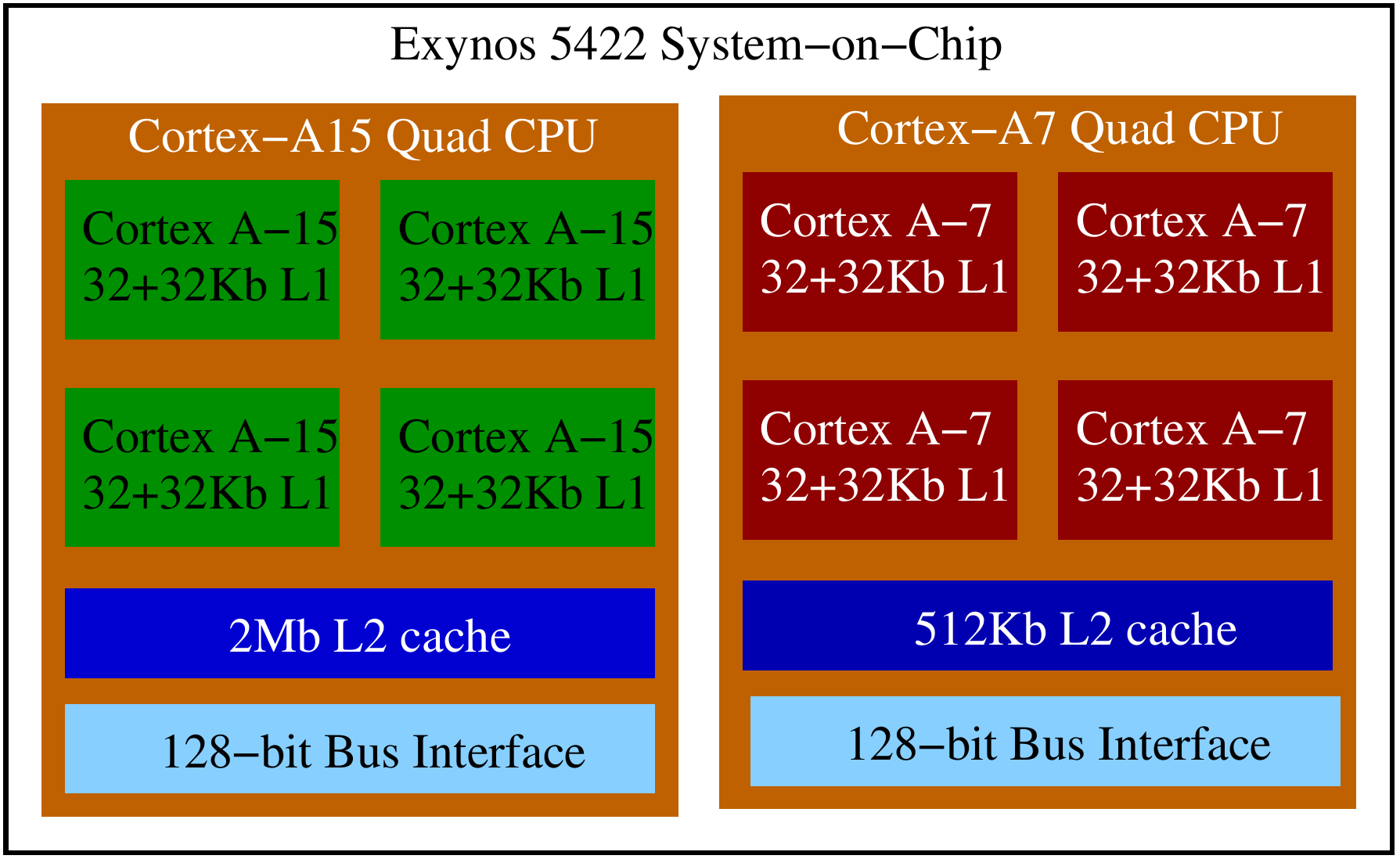}
\end{center}
\caption{\label{fig:exynos} Architecture block diagram for the Exynos 5422 SoC.}
\end{figure}

The ODROID-XU3 board employed in our experiments comprises a Samsung
Exynos 5422 SoC with an ARM Cortex-A15 quad-core processing cluster,
running at 2.0~GHz, and a Cortex-A7 quad-core processing
cluster, at 1.4~GHz. Both clusters access a shared DDR3 RAM
(2 Gbytes) via 128-bit coherent bus interfaces. Each ARM core (either
a Cortex-A15 or a Cortex-A7) has a 32+32-Kbyte L1 (instruction+data) cache.
In addition, the four ARM Cortex-A15 cores share a 2-Mbyte L2 cache, while the four
ARM Cortex-A7 cores share a smaller 512-Kbyte L2 cache; see Figure \ref{fig:exynos}.

The target platform runs an Ubuntu 14.04 LTS distribution with the Linux kernel 3.10. The library
was compiled with \texttt{-O3} optimization level of the \texttt{g++ 4.8.2} compiler. 
We ensure that all cores run at
their highest frequency during the experiments by setting 
the appropriate frequency limits
in the Linux performance governor. The codes are instrumented with the \texttt{pmlib}~\cite{AlonsoICPP12}
framework, which collects power dissipation samples corresponding to instantaneous
power readings from four independent sensors in the board (for the
Cortex-A7 cores, Cortex-A15 cores, DRAM DIMMs, and NEON SIMD engine), with a sampling rate
of 250~ms.
The power readings from all four sensors are added to estimate the instantaneous total power dissipation, and
this collection of values are averaged and multiplied by the execution time to obtain the energy consumption.

\subsection{Performance of the Multi-threaded libde265 on the ODROID-XU3}
\label{deco_initial}

To characterize the performance of the libde265 library, the developers  
show that, when decoding a 1080p HEVC bit-stream via WPP multithreading, 
on average the gstreamer plugin is able to process 150 FPS on a server equipped with a recent Intel 
desktop quad-core CPU. However, as we will show next, these encoding ratios are much lower for the 
ODROID-XU3 board.

The test set for our experiments includes the five 1080p videos from the JCT-VC benchmark~\cite{JCTVC-L1100};
hereafter, we report average results for all these video sequences.
These videos were previously encoded with the HM-16.2 reference encoder \cite{HM} using four QP points (22, 27, 32 and 37). 
The HM encoding parameters
were those set by default in the random-access 
main configuration, except in that 
the WPP option is enabled. 

\begin{figure}[t]
\begin{center}
\begin{tabular}{c}
\begin{minipage}[c]{\columnwidth}
\includegraphics[width=\textwidth]{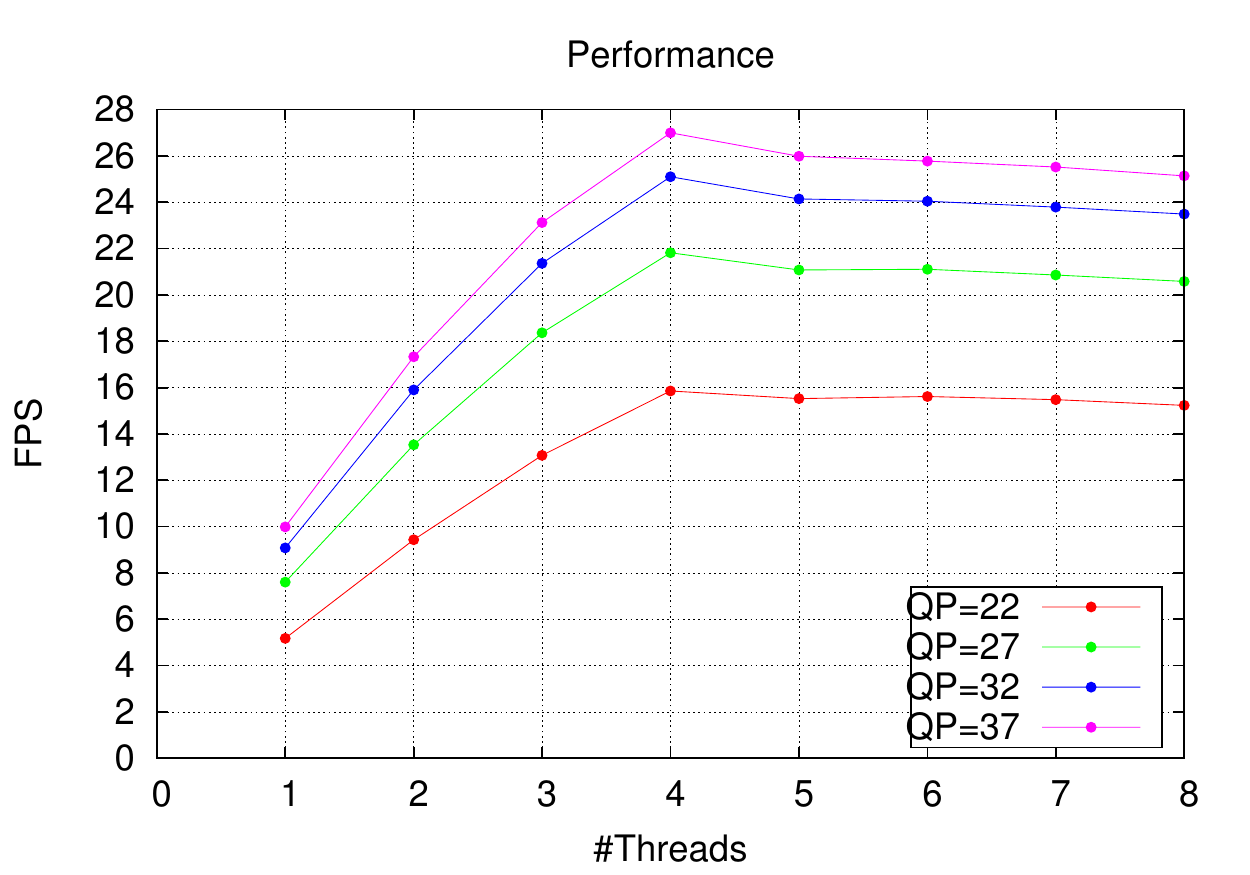}
\includegraphics[width=\textwidth]{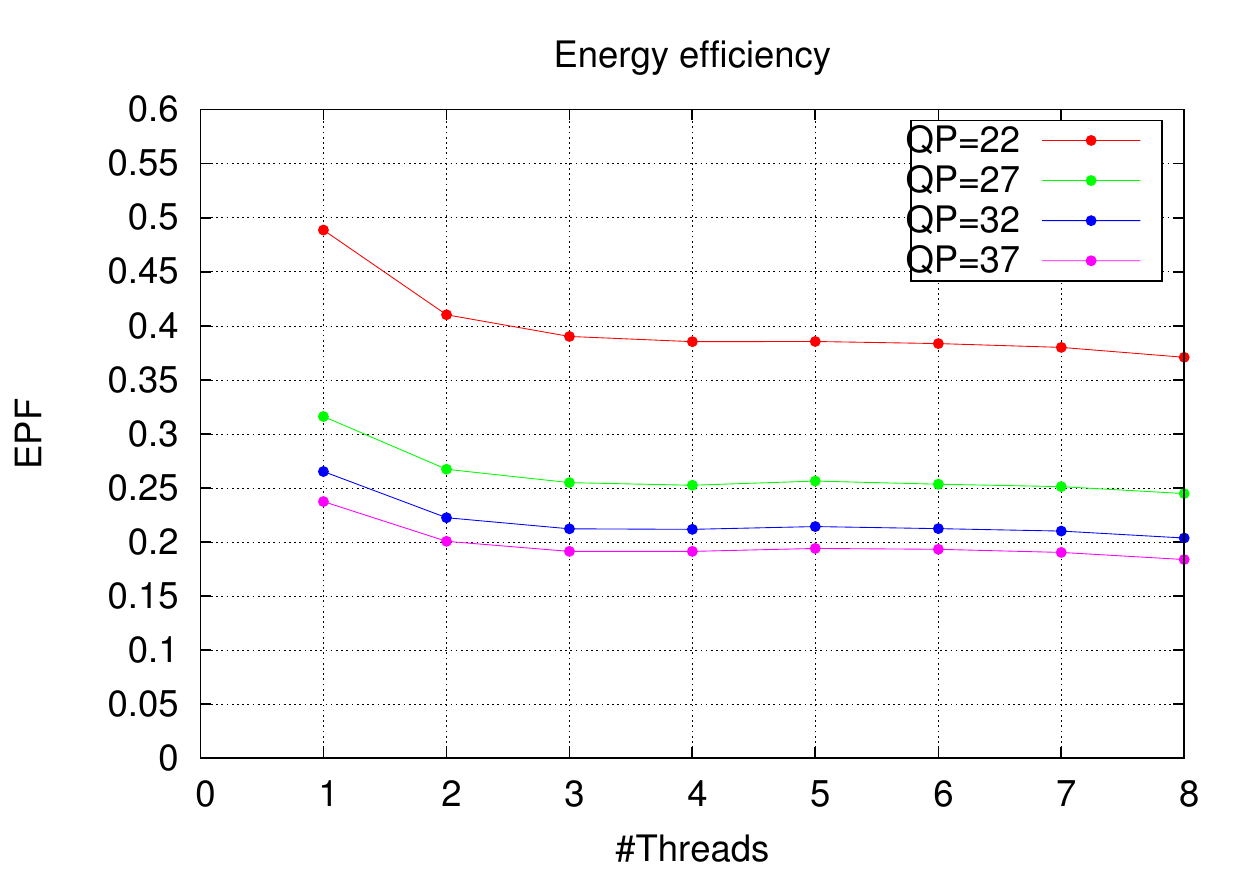}
\end{minipage}
\end{tabular}
\end{center}
\caption{Average FPS (top) and energy per frame (bottom) of the reference libde265 decoder on the ODROID-XU3.} 
\label{fig:performance_inicial}
\end{figure}

The top plot in~\figurename~\ref{fig:performance_inicial} shows the FPS (averaged for 
all five 1080p videos from the JCT-VC test set) on the ODROID-XU3. The bottom plot in 
the same figure shows the (average) energy (in Joules) per frame (EPF) consumed during these experiments. The 
first aspect to observe is that the use of more than four threads produces an unexpected decrease of the FPS.
The reason for this behavior is the Operating System (OS) scheduler,
which simply schedules all the threads to the Cortex-A15 cores. 
In consequence, when more than four 
threads are spawned, some Cortex-A15 cores interleave the execution
of two threads and, for this particular application, a reduction in the FPS rate is observed.

An additional aspect to note is that the FPS does not scale linearly with the number of cores, due to the 
dependencies when decoding the CTU rows. Concretely, as the videos are encoded to support WPP, the decoding 
of a CTU row cannot start till 
a minimum of two CTUs have been completely reconstructed for the previous CTU row.
As a result, 
at the beginning of decoding each frame, all threads except for one will have to wait before they commence to reconstruct 
their assigned CTU rows. 
To further expose the effect of WPP decoding on a multi-threaded execution,
let assume that the decoder spawns four (worker) threads. This means 
that the decoding process of the fourth CTU row cannot start until a minimum of six CTUs of the first CTU row have been reconstructed, 
which represents one fifth of the horizontal resolution for a 1080p video. 
Moreover, when eight threads are spawned, the reconstruction of the eighth CTU row 
cannot begin until a minimum of fourteen CTUs of the first CTU row have been reconstructed, 
which is almost half of the horizontal resolution of a 1080p video. 
In summary, with this approach, the level of concurrency is linearly reduced as the multithreading factor grows.
This explains why, in our experiments with QP=37, two threads deliver a speed-up of 1.82 over the sequential execution, 
but four threads offer a meager speed-up of 3.02.

Regarding the energy efficiency, using more than four threads does not increase significantly the energy efficiency of the 
HEVC decoder. In addition, four threads offer better energy efficiency than one, but the effect is less visible due 
to the sublinear scalability of the HEVC decoder.

\subsection{Static manual thread mapping on the Exynos 5422 SoC}
\label{deco_forced}

\begin{figure}[t]
\begin{center}
\begin{tabular}{c}
\begin{minipage}[c]{\columnwidth}
\includegraphics[width=\textwidth]{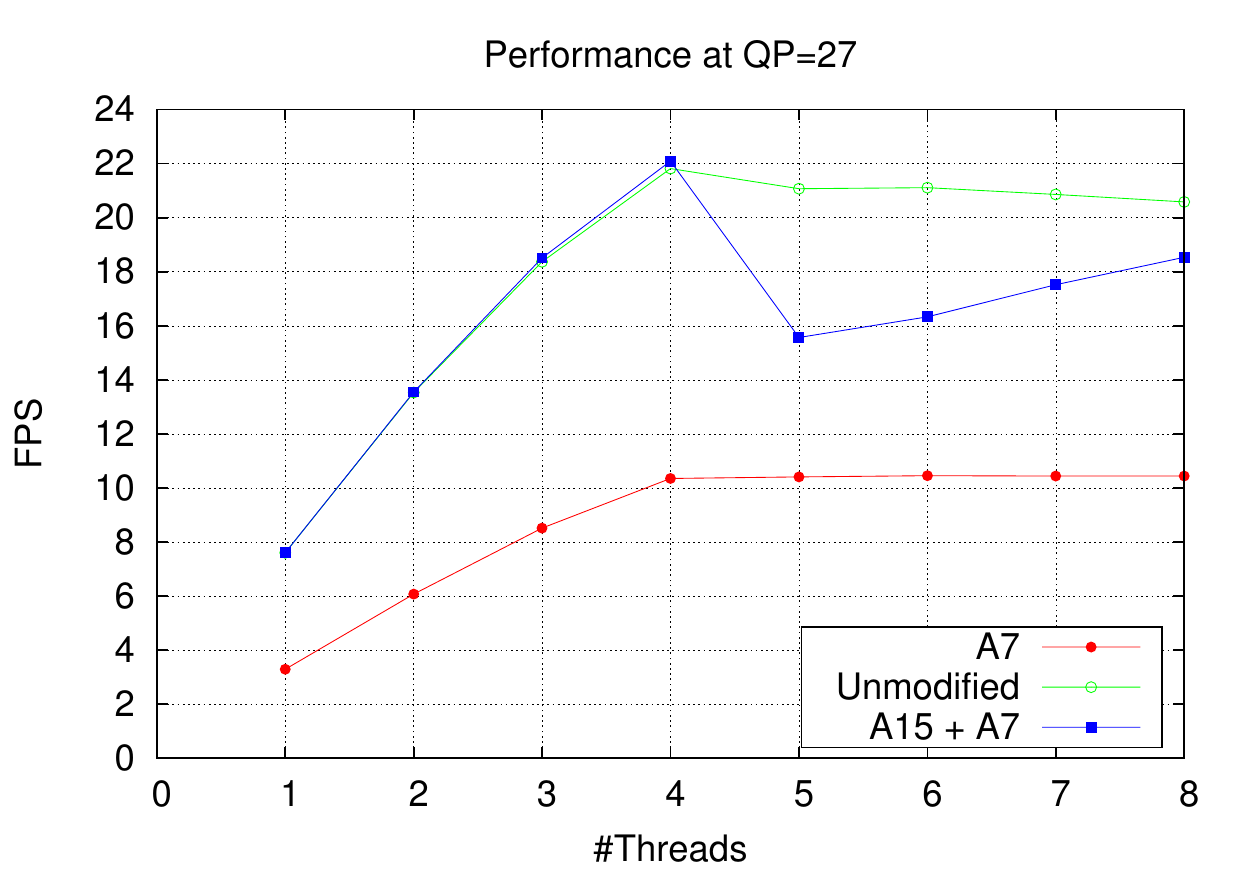}
\includegraphics[width=\textwidth]{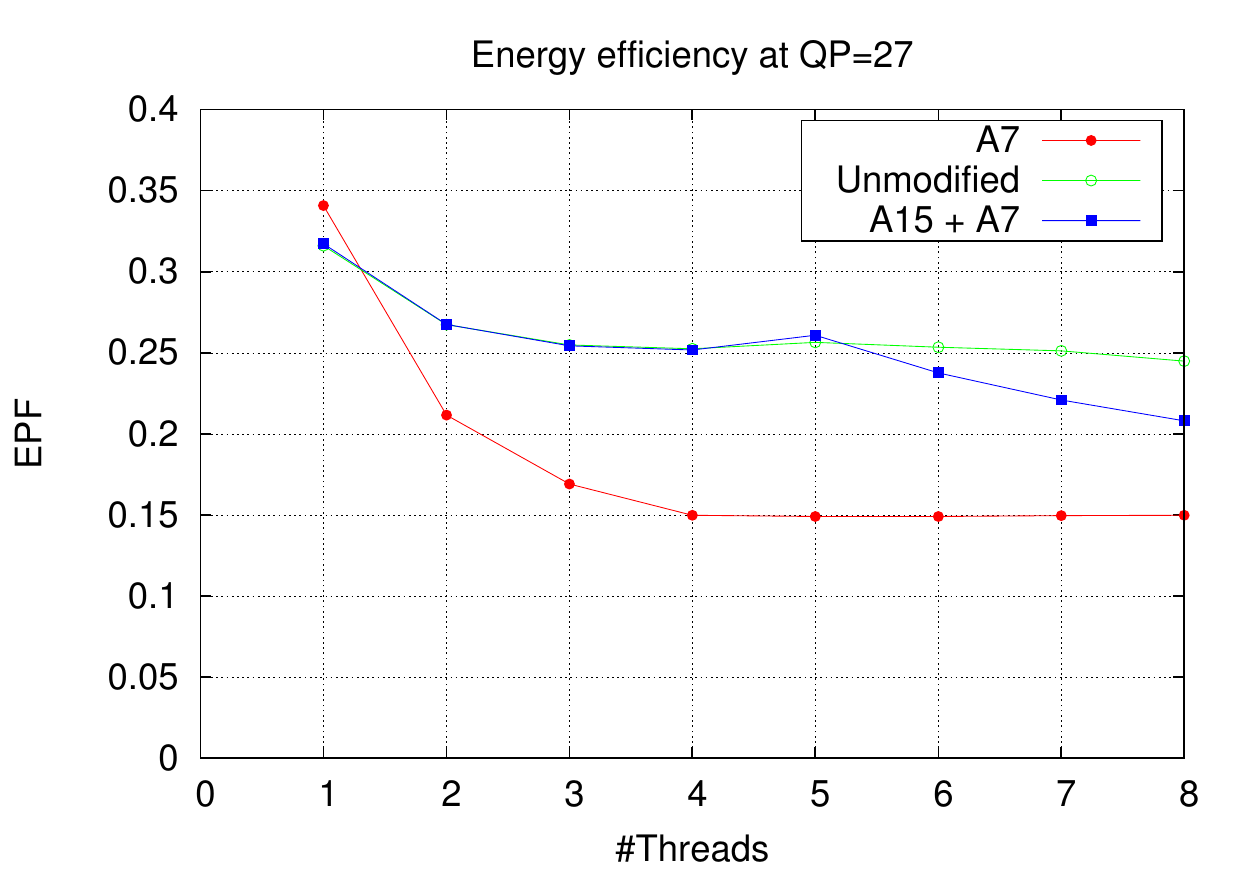}
\end{minipage}
\end{tabular}
\end{center}
\caption{Average FPS (top) and energy per frame (bottom) of the reference libde265 decoder on the ODROID-XU3 using 
         one type of core only (\textsf{A7} for the Cortex-A7 and \textsf{Unmodified} for the Cortex-A15), 
         and both types of cores via static manual binding (\textsf{A15 + A7}).}
\label{fig:performance_forced}
\end{figure} 

Our initial attempt to exploit the computational resources of the Exynos 5422 SoC more efficiently 
manually binds the (worker) thre\-ads to concrete cores, 
populating first each Cortex-A15 core with a single thread, and from then on mapping threads to
the Cortex-A7 cores. 

The two plots in~\figurename~\ref{fig:performance_forced} display the average FPS and EPF (top and bottom, respectively)
attained with the static manual binding (lines labelled as \textsf{A15 + A7}).
Additionally, for comparison purposes, we include the performance lines when the execution is 
limited to the Cortex-A7 cores (label \textsf{A7}), as well as the lines of the 
``\textsf{unmodified}'' version of the reference library that uses only the Cortex-A15 cores (see previous subsection). 
In all cases, for clarity, we only include the lines for a concrete QP, but the qualitative conclusions that can be 
extracted from all other QPs are similar. 

Let us focus on the FPS first. As expected, the results obtained with the static manual biding upon spawning four or less threads 
reveal performance rates that are very close to those observed with the unmodified version of the library.
The reason is that, in both configurations, we are only using the Cortex-A15 cores
(with static manual biding, threads are first mapped to this type of core; with the unmodified configuration, threads are only
mapped to them).
However, when five threads are used, the performance attained with the static manual binding is
significantly reduced, steadily growing from that point with the number of threads but remaining below the 
performance obtained with the unmodified version of the library. 
The source of this behaviour is that, due to the dependencies during the reconstruction of the CTU rows, 
the Cortex-A7 cores slow down the threads running on the Cortex-A15 cores, 
so that the latter basically proceed at the speed of the former. To illustrate this, note that the static manual biding scheme, operating
with the full 8~cores, delivers around 18.5~FPS, which is slighly below twice the FPS rate attained with 4~Cortex-A7 cores 
(around 10.1~FPS).

\begin{figure}[t]
\begin{center}
\begin{tabular}{c}
\begin{minipage}[c]{\columnwidth}
\includegraphics[width=\textwidth]{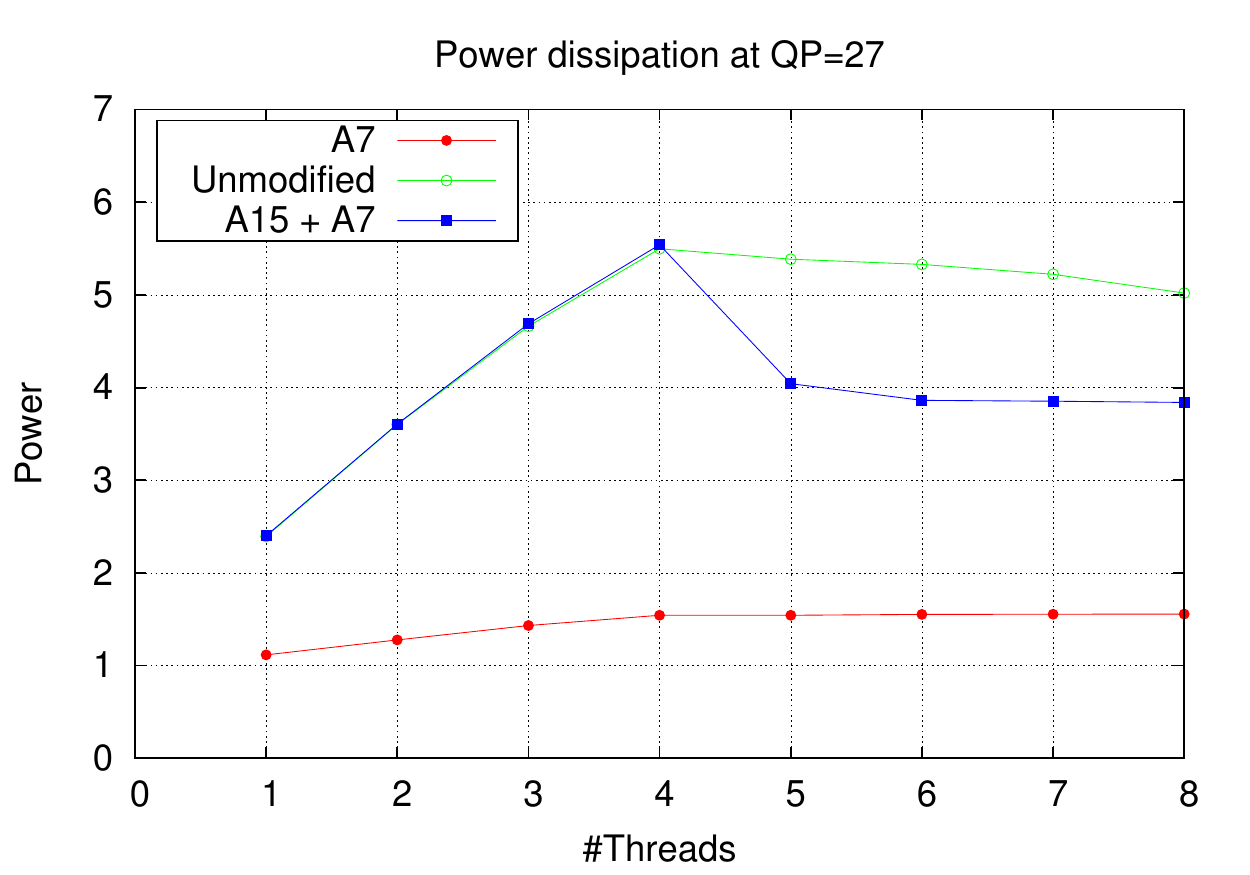}
\end{minipage}
\end{tabular}
\end{center}
\caption{Average power dissipation of the reference libde265 decoder on the ODROID-XU3 using 
         one type of core only (\textsf{A7} for the Cortex-A7 and \textsf{Unmodified} for the Cortex-A15);
         and both types of cores via static manual binding (\textsf{A15 + A7}).}
\label{fig:power_forced}
\end{figure} 

From the point of view of EPF, the static manual binding delivers higher energy efficiency 
than the unmodified version when five or more threads are spawned. This is explained in 
\figurename~\ref{fig:power_forced}, which displays the power dissipation (in Watts) of the different configurations.
There, we can observe the large differences in power consumption when 
the Cortex-A7 cores are only used (around 1.5 Watts) and when the unmodified version is used (around 5.5 Watts). 
We can also note that, when we exploit both types of cores
simultaneously via static manual binding, the average power draft decreases. 
In this scenario, the Cortex-A7 cores, which dissipate significantly less power than the Cortex-A15 cores, 
are at full load, but the Cortex-A15 cores are not, reducing the overall power rate.
A decrease of about 25-30\% in power dissipation 
is delivered when both types of cores are used simultaneously in comparison with the unmodified version, which is
more than the FPS decay, which ranges from 15\% to 28\%. The effect on the energy consumption
is explained by the linear dependence of this metric on the product of time and power.


In conclusion, these experiments naturally motivate the need for an architecture-aware alternative to the original 
multithreaded implementation, which is able to efficiently exploit the asymmetric resources of
the ARM big.LITTLE architecture (or any other AMP).  Ideally, 
an appropriate scheduling mechanism that exploits both the Cortex-A15 and Cortex-A7 cores should render two positive effects:
\begin{itemize}
\item An increment of FPS. To satisfy this, the Cortex-A7 cores should not slow down the execution of the Cortex-A15 cores but
positively contribute to the global (combined) decoding rate.

\item A decay of EPF. The Cortex-A7 cores are more energy-efficient than its Cortex-A15 counterparts, 
so by including the former we should save energy.
\end{itemize}

\section{Architecture-Aware Optimization of libde265 on the ARM big.LITTLE AMP}
\label{proposal}

The default (unmodified) approach adopted by libde265 on a multi-threaded CPU presents three major 
drawbacks when WPP is applied to simultaneously leverage all the cores of 
an ARM big.LITTLE AMP:
\begin{itemize}
\item Due to the differences in performance between the Cortex-A7 and Cortex-A15 cores,
in principle the OS scheduler does not map threads to the former.

\item When forcing the scheduler to use the Cortex-A7 cores, the dependencies intrinsic to
WPP do not allow to exploit the full potential of the Cortex-A15 cores, decreasing the overall FPS.

\item For WPP, the performance does not scale linearly with the number 
of threads, which directly affects the energy efficiency of the solution. This issue 
is not specific of an AMP but rather affects to any multi-threaded CPU. 
In the literature there exist several strategies that aim to mitigate this effect~\cite{chi12,chi13}.

\end{itemize} 
In the remainder of this section, we address the first two issues in order to 
improve the performance of the libde265 on the Exynos 5422 SoC. We do not consider the third issue
as it has been previously investigated in the literature.

\subsection{Asymmetry- and criticality-aware scheduling on the Exynos 5422 SoC}
\label{load_algorithm}

Taking into account the results in Sections~\ref{deco_initial} and~\ref{deco_forced}, 
we have introduced significant modifications
in the original multi-threaded implementation of the libde265 library in order to migrate the
threads between the two types of cores available in the Exynos 5422 SoC at execution time. 
Note that the optimizations exposed in this section are only beneficial when the execution proceeds on
a number of threads that exceeds the amount of Cortex-A15 cores,
as otherwise our solution only exploits the more powerful Cortex-A15.

\begin{figure}[t]
\begin{center}
\begin{tabular}{c}
\begin{minipage}[c]{0.9\columnwidth}
\includegraphics[width=\textwidth]{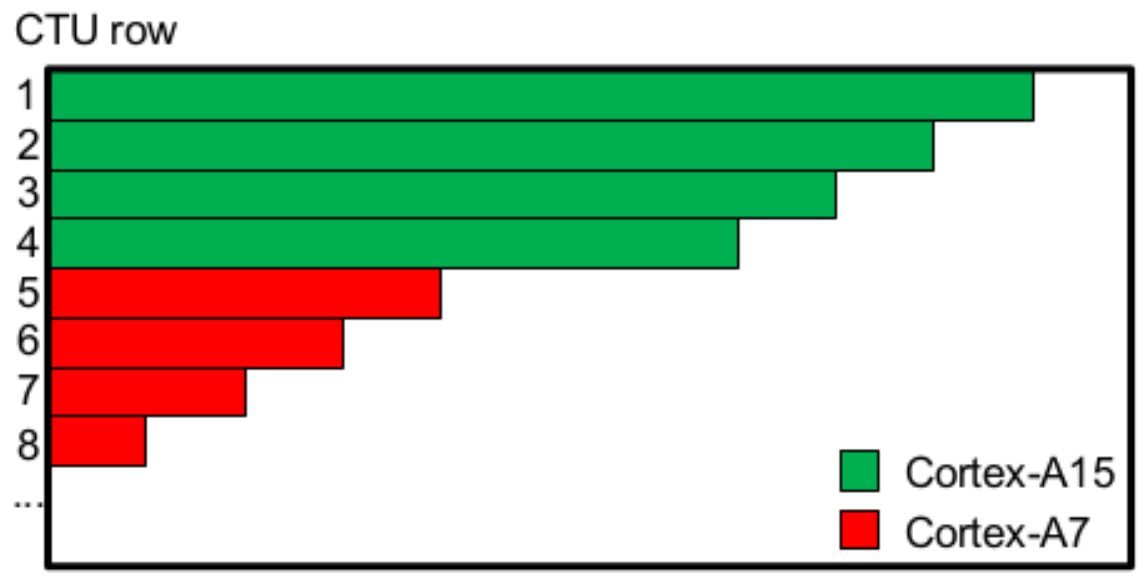}
\end{minipage}
\end{tabular}
\end{center}
\caption{Initial mapping of CTU rows to hardware cores.}
\label{fig:CTB_row_initial}
\end{figure}

The fundamental step towards delivering high performance and energy efficiency on the target 
Exynos 5422 SoC 
is to carry out an appropriate dynamic binding of the threads to the cores. For simplicity, let us assume an execution
with 8~threads. Then, at the beginning of 
decoding each frame, the four threads assigned to reconstruct the top four CTU rows are initially bound to 
Cortex-A15 cores, while the next four CTU rows are reconstructed by four threads that are bound 
to Cortex-A7 cores. 
This situation is graphically illustrated in \figurename~\ref{fig:CTB_row_initial}.

This initial mapping of threads to cores does not differ from the static manual binding previously presented.
The key to our new \textit{asymmetry-aware} solution, however, 
is to control the migration of threads, in order to ensure that the Cortex-A15 remain
in charge of the ``critical'' CTU rows.

\begin{figure}[t]
\begin{center}
\begin{tabular}{c}
\begin{minipage}[c]{\columnwidth}
\includegraphics[width=\textwidth]{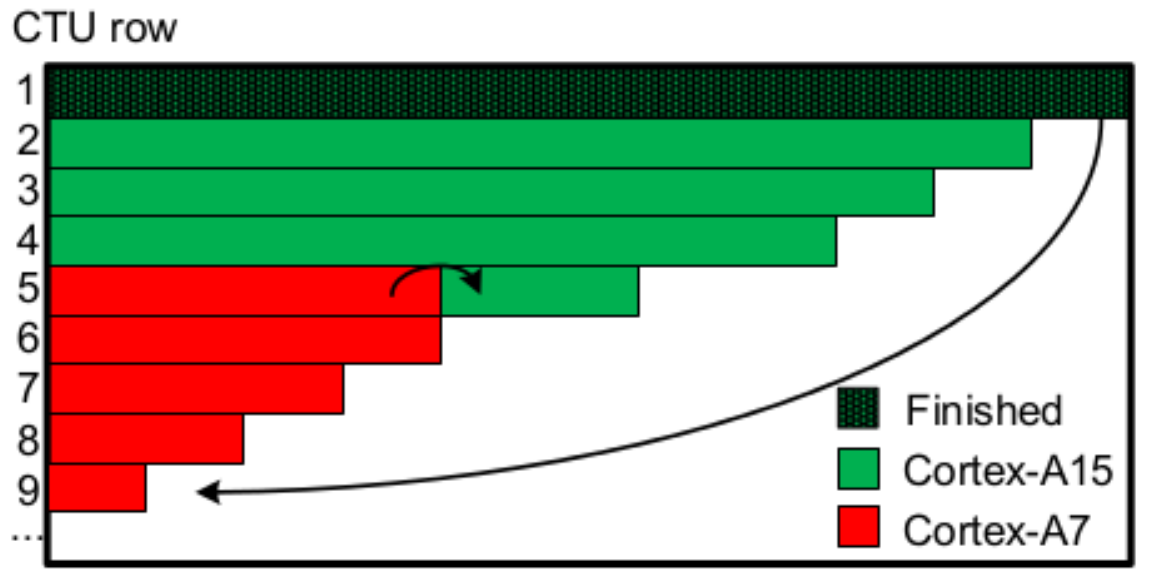}
\end{minipage}
\end{tabular}
\end{center}
\caption{Migration of task/thread between Cortex-A7 and Cortex-A15.}
\label{fig:CTB_row_change}
\end{figure}

Let us consider, for example, that 8~threads, denoted as {\sf T1}--{\sf T8}, are spawned, and they
commence to process
the top 8~CTU rows, with the {\sf i}-th row assigned to thread {\sf Ti}, the first four threads mapped
to the four Cortex-A15 cores ({\sf big.\{A,B,C,D\}}), and the next four to the four Cortex-A7 ({\sf LITTLE.\{A,B,C,D\}}).
Consider next that, starting from the initial scenario, 
thread {\sf T1}, which processes the first CTU row on {\sf big.A}, completes this task. 
In this scenario, thread {\sf T5}, in charge of the fifth CTU row on {\sf LITTLE.A}, is migrated to {\sf big.A},
where it continues processing the same row. In addition, thread {\sf T1} is migrated to 
{\sf LITTLE.A}, where it commences to process the 9th CTU row. 
As mentioned, 
the purpose here is to keep all 8~cores/threads busy with work, but to ensure that, from the 8~CTU rows that are (most of the
time) on-the-fly,
the top four are assigned to threads running on Cortex-A15 cores, while the remaining four are processed by threads
mapped to the Cortex-A7.  A graphical illustration of the migration is provided in \figurename~\ref{fig:CTB_row_change}.

From the implementation point of view, the threads
bound to the Cortex-A7 cores are responsible for checking if any Cortex-A15 core becomes ``idle''. To avoid incurring
an excessive overhead, this test is done every time a CTU is completely 
reconstructed. As a result, the migration of the threads from slow to fast cores is not immediate,
but can be slightly delayed.
Due to this, a special situation may occur that has to be tackled with care.
Assume that the first thread running on a Cortex-A7 that completes a CTU and, therefore, 
checks if there are any Cortex-A15 cores available is {\sf T6}, running on {\sf LITTLE.B}.
In this scenario, we only allow {\sf T6} to migrate to a vacant Cortex-A15 core if there are at least two
of them idle as, otherwise, we would be giving higher priority to the reconstruction of the 6th CTU row over the 5th one. 
Similarly, we will allow {\sf T7}/{\sf T8} to migrate to Cortex-A15 cores if at least three or all of them are idle.
With this strategy,
we ensure that the threads responsible for reconstructing the four pending CTU rows with highest priority 
(i.e, those in the top) are processed in the Cortex-A15 cores;
but we simultaneously allow that, when there are Cortex-A15 cores available, 
those threads processing CTU rows in the Cortex-A7 cores do not have to wait until the top one is migrated. 

In the final part of the CTU row reconstruction, we ensure that the threads which process the 
bottom CTU rows of each frame are never migrated to the Cortex-A7 cores, but 
remain bound to Cortex-A15 cores. 
As a consequence, 
in order to start with the in-loop filters part, we have four threads bound to Cortex-A15 cores and four 
bound to Cortex-A7 cores.

The multithreaded parallelization of the stage that applies the filters is easier as,
in this case,
there exist a larger volume of tasks and considerably higher level of concurrency (i.e.,
less dependencies) compared with the prior stage.
In consequence, thread scheduling during the application of the in-loop filters is not so crucial/complex.
The fundamental modification in the filter stage aims to keep the Cortex-A15 cores always busy with work. Thus, for example,
when there number of tasks available for execution is lower than eight, the 
Cortex-A7 cores are the ones that first become idle. 
This test is again carried out by the 
threads mapped to the Cortex-A7 cores whenever the filters are applied to a complete CTU.
If an idle Cortex-A15 core is detected, the thread that found this situation is migrated there.

\subsection{NEON intrinsics}
\label{NEON}

The libde265 library integrates SSE optimizations for x86-based architectures. Concretely, some functions of the library are 
implemented using SSE4 or SSE4.1 intrinsics. However, these instructions are not supported by ARM processors,
as these systems have their own SIMD instruction set, called NEON. In consequence, 
in order to exploit the SIMD units in the ARM architecture, it is necessary to transform 
the SSE-based functions to their NEON-based counterparts.
For this purpose, we have extended the work in~\cite{SSEtoNEON}, which 
proposes NEON equivalents for a subset of the SEE intrinsics. Following this path, we have completed 
that approach to deliver NEON-equivalent intrinsics for all SSE intrinsics appearing in the libde265 library.

\section{Performance evaluation}
\label{performance}

\begin{figure}[t]
\begin{center}
\begin{tabular}{c}
\begin{minipage}[c]{\columnwidth}
\includegraphics[width=\textwidth]{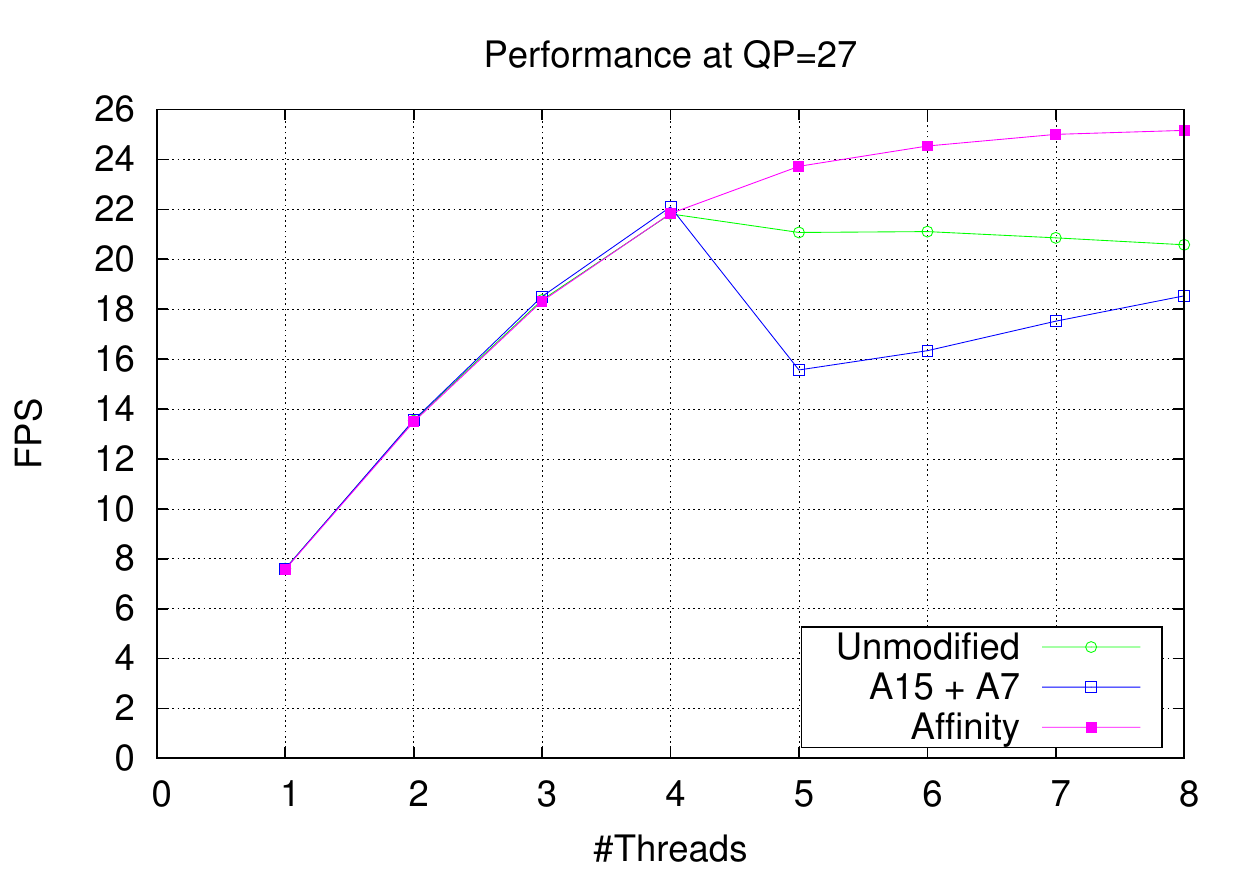}
\includegraphics[width=\textwidth]{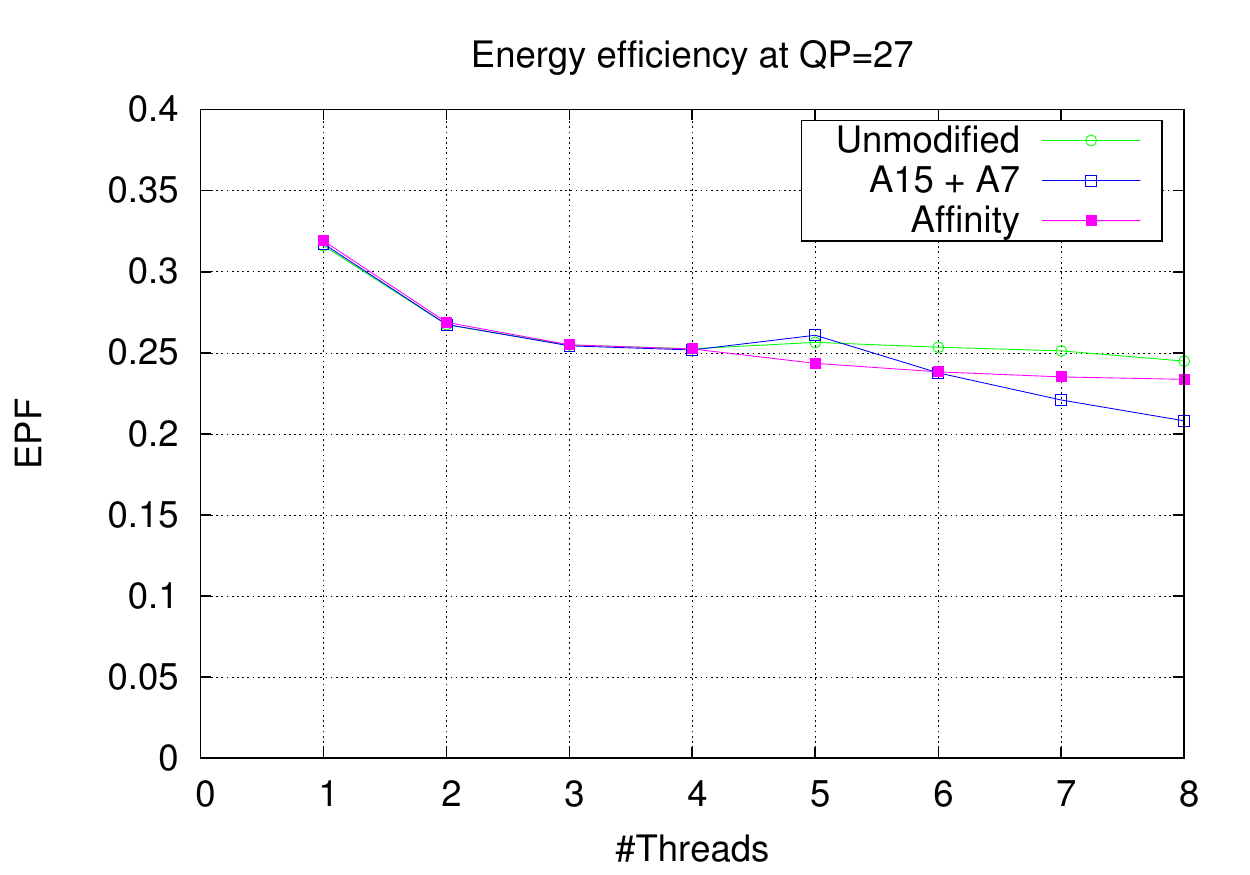}
\end{minipage}
\end{tabular}
\end{center}
\caption{Average FPS (top) and energy per frame (bottom) of the reference libde265 decoder on the ODROID-XU3 using
         only the Cortex-A15 cores (\textsf{Unmodified}); and
         both types of cores via manual binding without and with thread migration 
         (\textsf{A15 + A7} and \textsf{Affinity}, respectively).}

\label{fig:performance_affi}
\end{figure}

We next evaluate the performance of the asymmetry-aware version of the libde265 library, in terms of both 
FPS and EPF. To analyze the impact of the
new schedu\-ling strategy and the adoption of SIMD intrinsics, 
this section is divided into three parts. In the first subsection, we evaluate the asymme\-try-aware mechanism that migrates threads 
to keep the Cortex-A15 busy most of the time;
in the second subsection, we study the impact of integrating NEON intrinsics;
and in the final subsection, we summarize the results and analyze all the modifications
jointly.

\subsection{Asymmetry- and criticality-aware scheduling}

The plots in~\figurename~\ref{fig:performance_affi} 
display the average FPS (top)
and EFP (bottom)
of the asymmetry-aware scheduling with thread migration described in Section~\ref{load_algorithm}
(lines labelled with \textsf{Affinity}).
For comparison purposes, the plots also include
the results attained by the unmodified version of the library, as well as those
of the version where the threads are manually bound upon initialization to a concrete core
but no migration is allowed (labelled as \textsf{A15 + A7}).

Regarding the FPS rate, this evaluation shows significant benefits when the threads 
are migrated to keep the ``critical'' tasks (i.e., the top CTU rows) running on the Cortex-A15 cores. With this configuration, 
the Cortex-A7 cores do not slow down the Cortex-A15 cores but instead contribute to 
the FPS when the number of threads 
exceeds the amount of Cortex-A15 cores. We can still observe a non-linear increment of the FPS as, due to the dependencies implicit
in WPP,
the threads sometimes have to synchronize before they can reconstruct the assigned CTU rows; in addition, there is a certain 
overhead due to the differences in the performance between both types of cores.

\begin{figure}[t]
\begin{center}
\begin{tabular}{c}
\begin{minipage}[c]{\columnwidth}
\includegraphics[width=\textwidth]{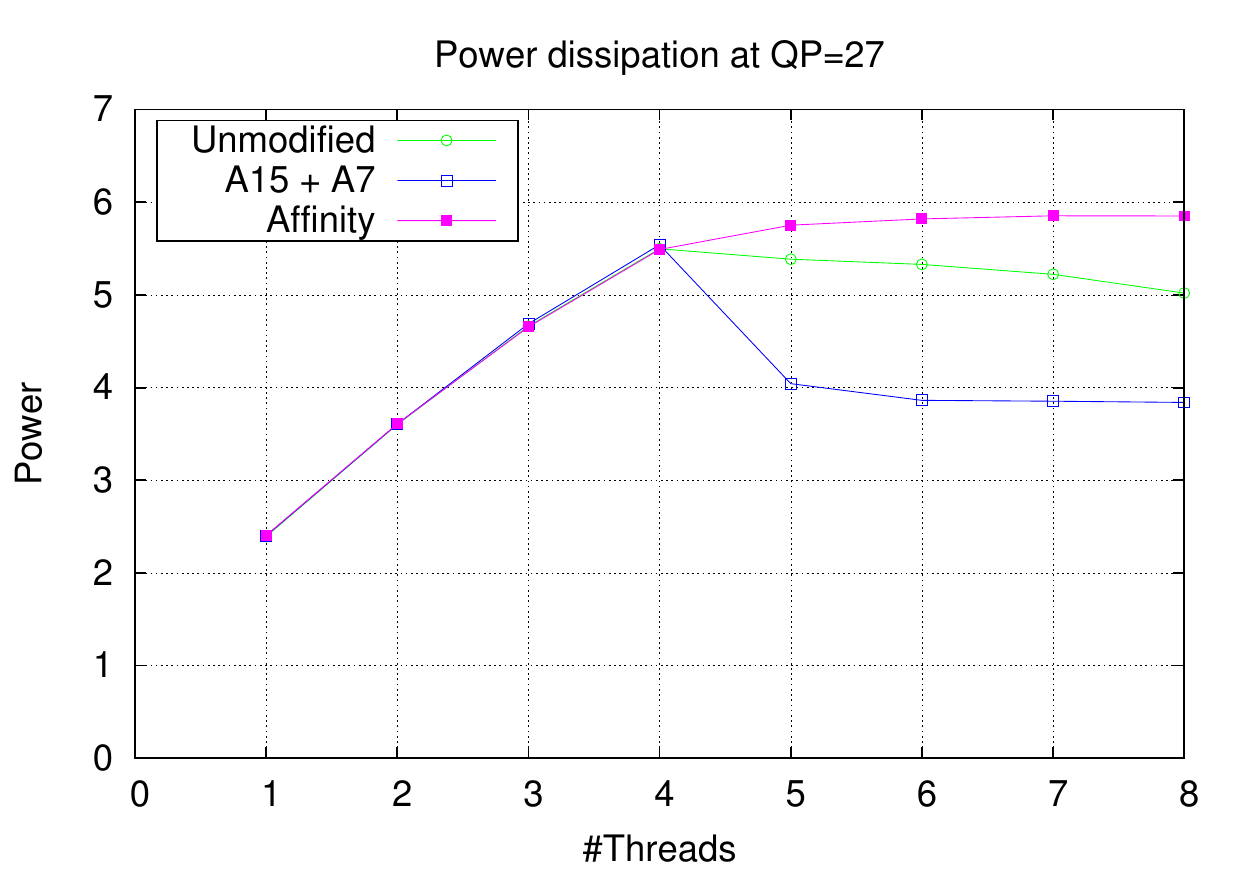}
\end{minipage}
\end{tabular}
\end{center}
\caption{Average power dissipation of the reference libde265 decoder on the ODROID-XU3 using
         only the Cortex-A15 cores (\textsf{Unmodified}); and
         both types of cores via manual binding without and with thread migration 
         (\textsf{A15 + A7} and \textsf{Affinity}, respectively).}
\label{fig:power_affi}
\end{figure} 

From the perspective of the EPF metric, the solution that integrates the enhancements presented in Section~\ref{load_algorithm} 
outperforms the unmodified version of the library but, unfortunately, it is still less energy-efficient than the approach that 
statically binds the threads to a specific core (no thread migration).
The reason of this behavior is motivated in \figurename~\ref{fig:power_affi}, which shows that the power dissipation 
grows with the number of threads. With the version enhanced with thread migration, 
all threads/cores are at full load (or close to it), and the power draft augments more rapidly
with the number of threads. However, the reduction of execution time is not enough to compensate the growth of power dissipation, resulting in a net increase of
energy consumption.

\begin{figure}[t]
\begin{center}
\begin{tabular}{c}
\begin{minipage}[c]{\columnwidth}
\includegraphics[width=\textwidth]{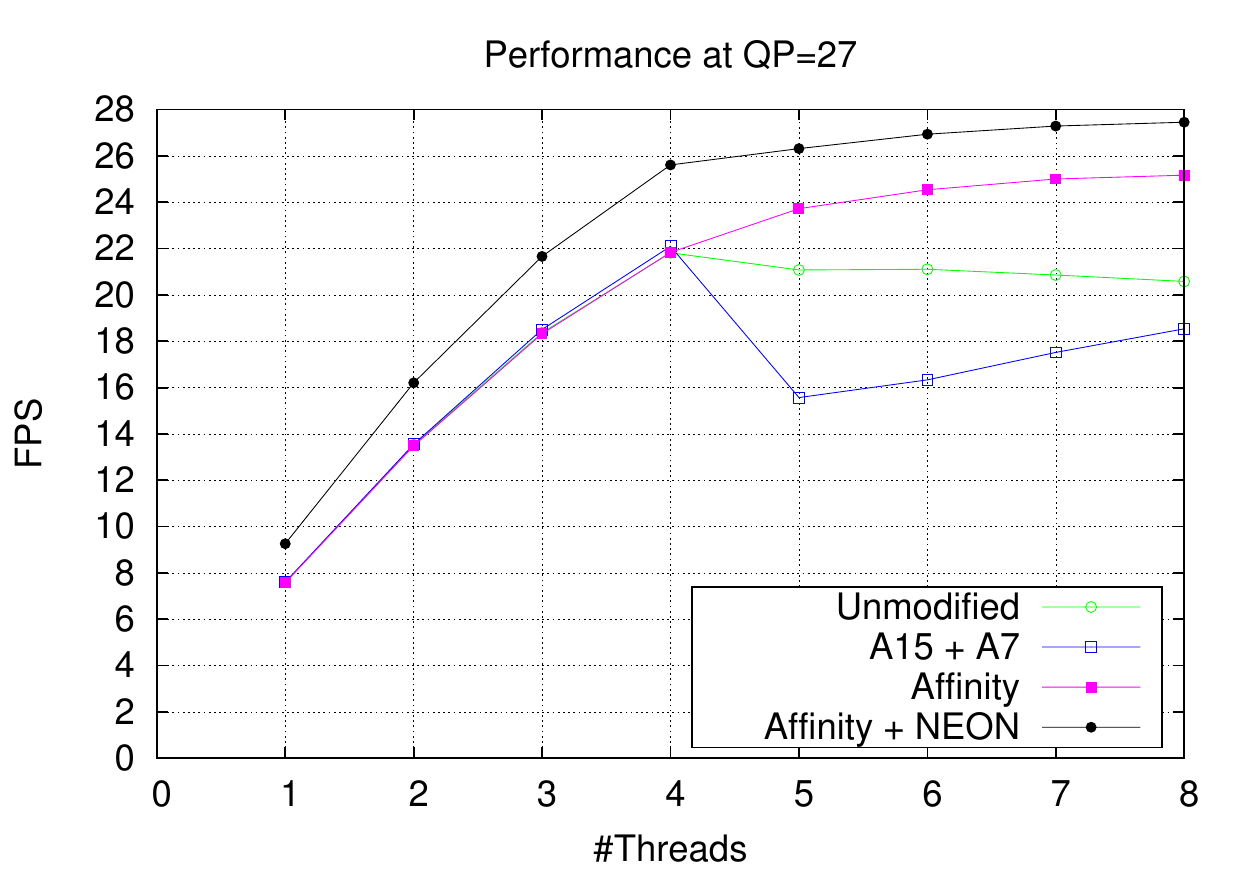}
\includegraphics[width=\textwidth]{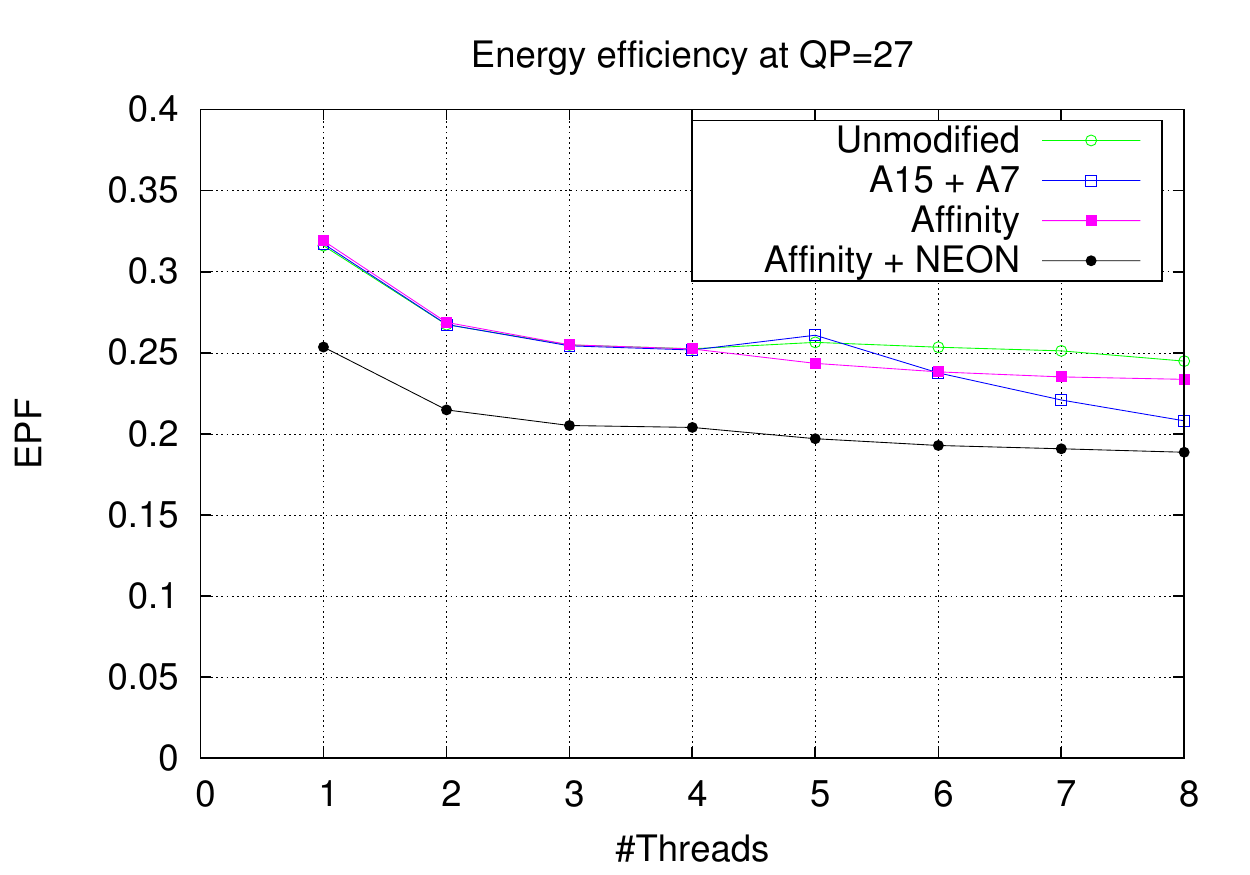}
\end{minipage}
\end{tabular}
\end{center}
\caption{Average FPS (top) and energy per frame (bottom) of the reference libde265 decoder on the ODROID-XU3 using 
         only the Cortex-A15 cores (\textsf{Unmodified});
         both types of cores via manual binding without and with thread migration 
         (\textsf{A15 + A7} and \textsf{Affinity}, respectively); and
         both types of cores via manual binding with thread migration 
         and NEON intrinsics (\textsf{Affinity + NEON}).}
\label{fig:performance_neon}
\end{figure}

\subsection{NEON intrinsics}

In principle, when using intrinsics in general, and NEON in the particular case of ARM cores, 
one can expect an increment in performance. However, an important aspect to analyze is the effect of this optimization
on energy consumption since, as we observed in the previous subsection, 
a faster execution does not necessarily imply a lower energy consumption. 

The plots in \figurename~\ref{fig:performance_neon} report the average FPS and EPF (top and bottom, respectively)
including now the version of the library that integrates the NEON 
optimizations with the asymmetry-aware scheduling that enforces thread migration when necessary 
(labelled as \textsf{Affinity + NEON}). 
For comparison purposes, we also show in the figure
the results of the most significant previous versions.

This figure reveals important benefits in terms of both FPS and EPF. The 
increase in the FPS rate was expected, since the positive effects of SIMD intrinsics 
have been widely exposed in the literature for numerical applications in general, and
for video encoders/decoders in particular. 

\begin{figure}[t]
\begin{center}
\begin{tabular}{c}
\begin{minipage}[c]{\columnwidth}
\includegraphics[width=\textwidth]{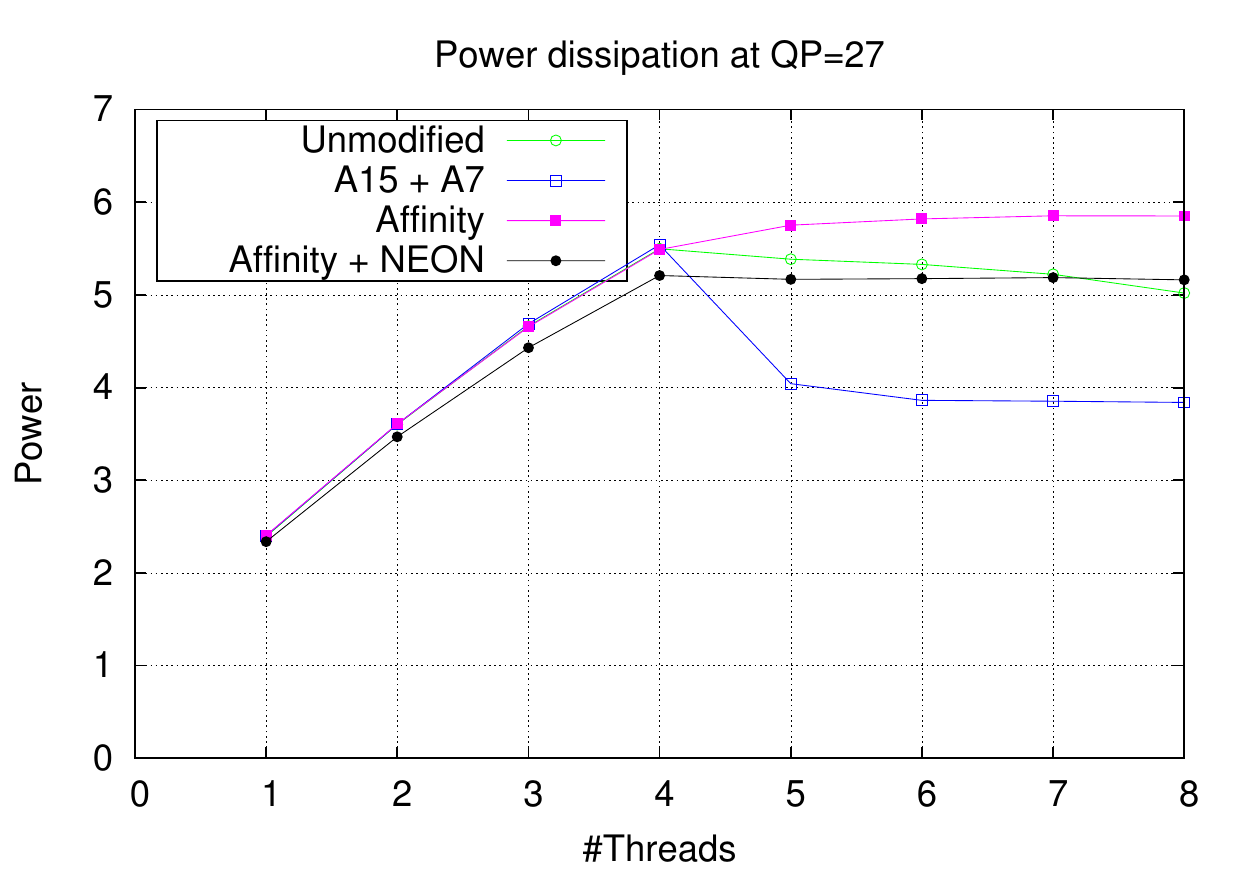}
\end{minipage}
\end{tabular}
\end{center}
\caption{Average power dissipation of the reference libde265 decoder on the ODROID-XU3 using 
         only the Cortex-A15 cores (\textsf{Unmodified});
         both types of cores via manual binding without and with thread migration 
         (\textsf{A15 + A7} and \textsf{Affinity}, respectively); and
         both types of cores via manual binding with thread migration 
         and NEON intrinsics (\textsf{Affinity + NEON}).}
\label{fig:power_neon}
\end{figure}

On the other hand, due to linear dependency 
between energy consumption and power-execution time, the reduction in EPF could be 
also expected if the power dissipation rate is not increased by a factor that exceeds the reduction of
execution time.
\figurename~\ref{fig:power_neon} shows the average power dissipation for the experiments presented in 
\figurename~\ref{fig:performance_neon}. There we can observe that the adoption of the NEON intrinsics yields a significant reduction 
of the average power dissipation compared to the asymmetry-aware version of the library without intrinsics.
Together with the reduction in execution time, this explains the notable gains in energy efficiency thanks to
the integration of the NEON intrinsics.

\subsection{Global comparison}

\begin{table*}[t]
\caption{Average FPS of the reference libde265 decoder on the ODROID-XU3 using 
         only the Cortex-A15 cores (\textsf{Unmodified});
         both types of cores via manual binding without and with thread migration 
         (\textsf{A15 + A7} and \textsf{Affinity}, respectively); and
         both types of cores via manual binding with thread migration 
         and NEON intrinsics (\textsf{Affinity + NEON}).}
\centering
\label{tab:1}       
\resizebox{\linewidth}{!}{
\begin{tabular}{|c||r|rr|rrr|rrr|}
\hline
\#Threads & \textsf{Unmodified} & \multicolumn{2}{c|}{\textsf{A15 + A7}} & \multicolumn{3}{c|}{\textsf{Affinity}} & \multicolumn{3}{c|}{\textsf{Affinity + NEON}}  \\
		&			 &			& \% vs 	 &	   & \% vs 		& \% vs    &	 & \% vs 	& \% vs  \\
		& FPS		 & FPS		& \textsf{Unmodified} & FPS & \textsf{Unmodified} & \textsf{A15 + A7} & FPS & \textsf{A15 + A7} & \textsf{Affinity}\\
\hline\hline
1 & 7.963 & 7.934 & -- & 7.916 & -- & -- & 9.844 & 24.07 & 24.35 \\
2 & 14.049 & 14.054 & -- & 13.972 & -- & -- & 17.028 & 21.16 & 21.87\\
3 & 18.982 & 19.052 & -- & 18.883 & -- & -- & 22.545 & 18.33 & 19.39\\
4 & 22.441 & 22.655 & -- & 22.375 & -- & -- & 26.401 & 16.54 & 17.99\\
5 & 21.682 & 16.102 & -25.73 & 24.220 & 11.71 & 50.42 & 26.909 & 67.12 & 11.11 \\
6 & 21.636 & 16.837 & -22.18 & 25.035 & 15.71 & 48.69 & 27.499 & 63.32 & 9.84\\
7 & 21.411 & 18.027 & -15.81 & 25.491 & 19.06 & 41.41 & 27.831 & 54.39 & 9.18\\
8 & 21.109 & 19.116 & -9.44 & 25.628 & 21.41 & 34.07 & 27.968 & 46.30 & 9.13\\
\hline
\end{tabular}
}
\end{table*}

\begin{table*}[t]
\caption{Average EPF of the reference libde265 decoder on the ODROID-XU3 using 
         only the Cortex-A15 cores (\textsf{Unmodified});
         both types of cores via manual binding without and with thread migration 
         (\textsf{A15 + A7} and \textsf{Affinity}, respectively); and
         both types of cores via manual binding with thread migration 
         and NEON intrinsics (\textsf{Affinity + NEON}).}
\centering
\label{tab:2}       
\resizebox{\linewidth}{!}{
\begin{tabular}{|c||r|rr|rrr|rrr|}
\hline
\#Threads & \textsf{Unmodified} & \multicolumn{2}{c|}{\textsf{A15 + A7}} & \multicolumn{3}{c|}{\textsf{Affinity}} & \multicolumn{3}{c|}{\textsf{Affinity + NEON}}  \\
		&			 &			& \% vs 	 &	   & \% vs 		& \% vs    &	 & \% vs 	& \% vs  \\
		& EPF		 & EPF		& \textsf{Unmodified} & EPF & \textsf{Unmodified} & \textsf{A15 + A7} & EPF & \textsf{A15 + A7} & \textsf{Affinity}\\
\hline\hline
1 & 0.327 & 0.325 & -- & 0.327 & -- & -- & 0.260 & -19.90 & -20.28\\
2 & 0.275 & 0.275 & -- & 0.276 & -- & -- & 0.221 & -19.38 & -19.67\\
3 & 0.262 & 0.262 & -- & 0.262 & -- & -- & 0.212 & -18.99 & -19.00 \\
4 & 0.260 & 0.259 & -- & 0.260 & -- & -- & 0.210 & -18.53 & -19.18\\
5 & 0.262 & 0.266 & 1.37 & 0.250 & -4.65 & -5.93 & 0.203 & -23.86 & -19.06\\
6 & 0.260 & 0.243 & -6.53 & 0.245 & -6.06 & 0.51 & 0.198 & -18.63 & -19.04\\
7 & 0.258 & 0.227 & -12.07 & 0.241 & -6.67 & 6.14 & 0.196 & -13.96 & -18.94\\
8 & 0.251 & 0.212 & -15.48 & 0.238 & -4.92 & 12.49 & 0.193 & -8.74 & -18.88\\
\hline
\end{tabular}
}
\end{table*}

\tablename~\ref{tab:1} and \tablename~\ref{tab:2} gather the 
performance of the distinct versions of the library in terms of FPS and EPF, respectively (averaged
for all JCT-VC videos and four QPs). 
Moreover, the tables quantify the differences (in \%) between these implementations, with
positive values reflecting an increment in the FPS/EPF, while negative values corresponding to a decrement.
Note that all versions 
which spawn less than five threads are equivalent to the unmodified implementation, except when the NEON intrinsics are exploited.

The results in the first table expose that, for this particular SoC, the unmodified version of the library is not able to deliver
the standard frame rate of 24 FPS for 1080p videos although it keeps four Cortex-A7 cores idle, which could
have been used for this purpose. 
A simple static-manual binding of the threads upon initialization, which does not
promote thread migrations, does not increase performance and, when the Cortex-A7 
cores are used,  the performance is even decreased by factors that range between 9\% and 25\% with respect to the default
(unmodified) configuration. 
Compared with this, the FPS rates are notably increased  
when the threads are migrated between the available cores while taking into account the criticality of the top CTU rows, 
as well as by leveraging the NEON units. 
For example, compared with the unmodified version, the thread migrating policy yields an
increase of performance between 11\% and 21\%;
and the integratation of NEON intrinsics
(four or less threads in the main column labeled as \textsf{Affinity + Neon})
delivers between 17\% and 24\% higher performance. 

From the perspective of energy efficiency, in the second table we observe that all versions reduce the EPF rate of the 
unmodified version of the library. However, despite the improvement of the FPS rate attained via
thread migration, this version is less energy-efficient than 
the one which carries out a static mapping of threads upon initialization. 
On the other hand, adding the NEON intrinsics is always beneficial, respectively delivering around 19\% and 9--24\% 
more energy efficiency 
than the asymmetry-aware and static mapping counterparts which do not exploit the NEON SIMD engine.

\section{Conclusions}
\label{conclu}

We have proposed and evaluated an asy\-mmetry-aware scheduling implementation of a reference
HEVC decoder for the ARM big.LITTLE AMP embedded in the Exynos 5422 SoC/ODROID-XU3 board. 
Our solution follows the parallelization approach
dictated by WPP to distribute the workload (CTU rows/tasks) among the fast (big) Cortex-A15 and the slow (LITTLE) Cortex-A7 cores
on-the-fly, migrating the threads in charge of executing those tasks with higher priority to
the former type of core. Moreover, the new implementation of the HEVC 
decoder is enhanced with NEON SIMD counterparts for all SSE intrinsics included in the reference
implementation of the library.

Our results reveal excellent improvements in performance compared with the execution of the architecture-oblivious 
reference implementation, which only exploits the big cores and cannot attain 1080p real-time decoding. 
In addition, we  demonstrate
that the exploitation of the Cortex-A7 cores not only enhances the overall performance, but also contributes
to improve the energy efficiency of decoding pipeline.

\begin{acknowledgements}
This work was supported by projects CICYT TIN2011-23283 and TIN2014-53495-R of
MINECO and FEDER. 
\end{acknowledgements}




%
%

\end{document}